\documentclass[prl,twocolumn,superscriptaddress]{revtex4-1}
\pdfoutput=1
\usepackage{amssymb}
\usepackage{amsfonts}
\usepackage{amsmath}
\usepackage{mathrsfs}
\usepackage[toc,page]{appendix}
\usepackage{hyperref}
\usepackage{graphicx}
\usepackage{enumerate}

\newcommand{\R}{\mathbb{R}}
\newcommand{\C}{\mathbb{C}}
\newcommand{\Z}{\mathbb{Z}}
\newcommand{\cS}{\mathbb{S}}
\newcommand{\mz}{\mathcal{Z}}
\newcommand{\mf}{\mathcal{F}}
\newcommand{\mn}{\mathcal{N}}
\newcommand{\sads}{\mathscr{S}}
\newcommand{\bx}{\bar{x}}
\newcommand{\ccc}{\cite{Russo:2016ueu}}

\begin{document}

\title[ On \textrm{S}$\mathrm{QED}_{3}$ and \textrm{S}$\mathrm{QCD}_{3}$ ]{%
On \textrm{S}$\mathrm{QED}_{3}$ and \textrm{S}$\mathrm{QCD}_{3}$: phase
transitions and integrability}
\author{Leonardo Santilli}
\email{lsantilli@fc.ul.pt}
\author{Miguel Tierz }
\email{tierz@fc.ul.pt}

\affiliation{Departamento de Matem\'{a}tica, Grupo de F\'{i}sica Matem\'{a}tica, Faculdade
de Ci\^{e}ncias, Universidade de Lisboa, Campo Grande, Edif\'{i}cio C6,
1749-016 Lisboa, Portugal.}

\begin{abstract}
We study supersymmetric Yang-Mills theories on the three-sphere, with
massive matter and Fayet-Iliopoulos parameter, showing second order phase
transitions for the non-Abelian theory, extending a previous result for the
Abelian theory. We study both partition functions and Wilson loops and also
discuss the case of different $R$-charges.
Two interpretations of the partition function as eigenfunctions of the $%
A_{1} $ and free $A_{N-1}$ hyperbolic Calogero-Moser integrable model are
given as well.
\end{abstract}

\maketitle



The study of supersymmetric gauge theories in curved
space-times has been pushed forward considerably in the last decade due
to the extension of the localization method of path integrals \cite{Pestun:2007rz,Kapustin:2009kz}. By using localization, a much
simpler integral representation of the observables of the gauge theories is achieved. In turn, these seemingly simple representations, in general
of the matrix model type, contain a wealth of information of different type.
First, they are very useful for asymptotic analysis and, in suitable large $N
$ double scaling limits, have predicted phase transitions in the theory \cite{Russo:2013sba,Russo:2014bda,Russo:2016ueu,Santilli:2018byi}. Secondly, in many cases,
especially for three dimensional theories, they are amenable to exact
analytical solutions, even for finite $N$ \cite{Russo:2014bda,Giasemidis:2015ial}. Such exact evaluation, or the procedure
leading to it, oftentimes may point towards a connection between the gauge
theory and, for example, integrable systems \cite{Teschner:2016yzf}.

All these aspects of the localization integral formulas will be exposed in
what follows, as we will not only study finite and large $N$ properties,
together with phase transitions in double scaling limits, but also give an
integrable systems view of the gauge theory, by showing a connection with
the hyperbolic Calogero-Moser system.

In what follows, we will consider $\mathcal{N}=4$ theory on the $3d$ sphere $%
\mathbb{S}^{3}$, with gauge group $U(n)$ and an even number $N_{f}=2N$ of massive chiral multiplets in the
fundamental, $N$ of them with mass $m$ and $N$ with mass $-m$, arranged into 
$N$ hypermultiplets. We also insert a Fayet-Iliopoulos (FI) term. Localization \cite{Kapustin:2009kz,Jafferis:2010un,Hama:2010av} gives the integral
representation of the partition function: 
\begin{align}
\mathcal{Z}_{N}^{U(n)}& =\int_{\mathbb{R}^{n}}d^{n}x\prod_{1\leq j<k\leq
n}\left( 2\sinh \frac{x_{j}-x_{k}}{2}\right) ^{2}  \notag  \label{eq:znc} \\
& \times \prod_{j=1}^{n}\frac{e^{i\eta x_{j}}}{2^{N}\left[ \cosh
(x_{j})+\cosh (m)\right] ^{N}},
\end{align}%
where we set the radius of $\cS^3$ to $1/2\pi $ and $\eta$ is the FI parameter. We will eventually
be interested in the limit in which the number of flavours $N_{f}=2N$ is
large, while the number of colours $n$ is kept finite. Therefore, we
consider $N_{f}=2N\geq 2n$, so that the integral \eqref{eq:znc} in
convergent, besides the theory is ``good'' (or ``ugly'', if $N=n$) according to the classification \cite{Gaiotto:2008ak}.

The Abelian case $n=1$ was studied in detail in \cite{Russo:2016ueu}. In
what follows, we will extend the results of \cite{Russo:2016ueu}, including $%
1/N$ corrections and the analysis of Wilson loops, as well as carrying over
the study to non-Abelian theories, $n>1$. In the simplest non-Abelian case $n=2$
we will also compute $1/N$ corrections to the large $N$ limit.

\paragraph*{Abelian theory at finite $N$.}

The partition function of the Abelian theory reads: 
\begin{equation}
\mathcal{Z}_{N}^{U(1)}=2^{-N}\int_{-\infty }^{+\infty }dx  e^{i\eta x}\left[
\cosh (x)+z\right] ^{-N},  \label{eq:z1}
\end{equation}%
where $z\equiv \cosh (m)$. The expression is significantly simpler than any
non-Abelian case, since the one-loop determinant of the vector multiplet is
trivial for $n=1$. The partition function \eqref{eq:z1} can be computed
exactly in terms of a hypergeometric function \cite{Russo:2016ueu}, as 
\begin{align}
\mathcal{Z}_{N}^{U(1)}& =\frac{\sqrt{2\pi }}{2^{N}(1+z)^{N-\frac{1}{2}}}%
\frac{\Gamma (N+i\eta )\Gamma (N-i\eta )}{\Gamma (N)\Gamma \left( N+\frac{1}{%
2}\right) }  \notag  \label{eq:exactz1} \\
& \times {_{2}F_{1}}\left( \frac{1}{2}-i\eta ,\frac{1}{2}+i\eta ,N+\frac{1}{2%
},\frac{1-z}{2}\right) .
\end{align}%
Using an Euler transformation for the hypergeometric \cite[Ch.2]{Schindler},
we can rewrite \eqref{eq:exactz1} when $\eta \geq 1,m\geq 1$ as: 
\begin{align}
\mathcal{Z}_{N}^{U(1)}& =\frac{e^{i\eta m}}{2^{N}(\sinh (m))^{N}}\frac{%
\Gamma (N-i\eta )\Gamma (i\eta )}{\Gamma (N)}  \notag \\
& \times {_{2}F_{1}}\left( 1-N,N,1-i\eta ,-(e^{2m}-1)^{-1}\right)   \notag \\
& +(\mathrm{replace}\ i\eta \longleftrightarrow -i\eta ).
\end{align}%
This latter form is illustrative: since the first coefficient, $a=1-N$, is a
nonpositive integer, the hypergeometric series terminates and gives a
polynomial of degree $N-1$ in the variable $y\equiv -(e^{2m}-1)^{-1}$.
Moreover, in our case the second coefficient $b=N=1-a$, thus the
hypergeometric function is actually an associated Legendre function of
imaginary order \footnote{%
Eq. 15.9.21 in \href{https://dlmf.nist.gov/15.9}{dlmf.nist.gov/15.9}.}: 
\begin{equation*}
{_{2}F_{1}}\left( 1-N,N,1-i\eta ,y\right) =\Gamma (1-i\eta )\left( \frac{y}{%
1-y}\right) ^{\frac{i\eta }{2}}P_{N-1}^{i\eta }(1-2y).
\end{equation*}%
The partition function reads: 
\begin{align*}
\mathcal{Z}_{N}^{U(1)}& =\frac{\pi e^{-\frac{\pi \eta }{2}}\Gamma (N-i\eta )%
}{2^{N}i\sinh (\pi \eta )\sinh (m)^{N}\Gamma (N)}P_{N-1}^{i\eta }(\coth (m))
\\
& +(\mathrm{replace}\ i\eta \longleftrightarrow -i\eta ),
\end{align*}%
where we used the property $\Gamma (1-i\eta )\Gamma (i\eta )=\pi /\sin (i\pi
\eta )$.

\begin{figure}[bth]
\centering
\includegraphics[width=0.4\textwidth]{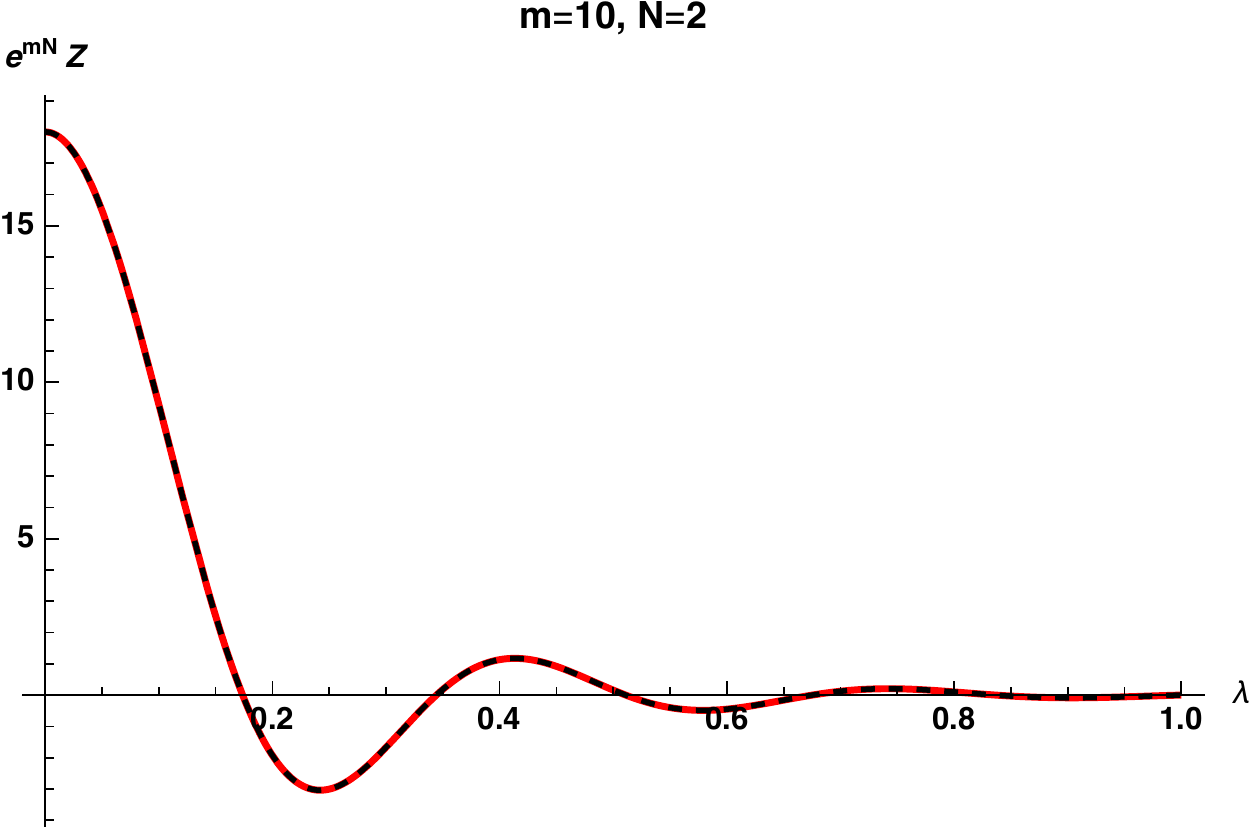}\newline
\includegraphics[width=0.4\textwidth]{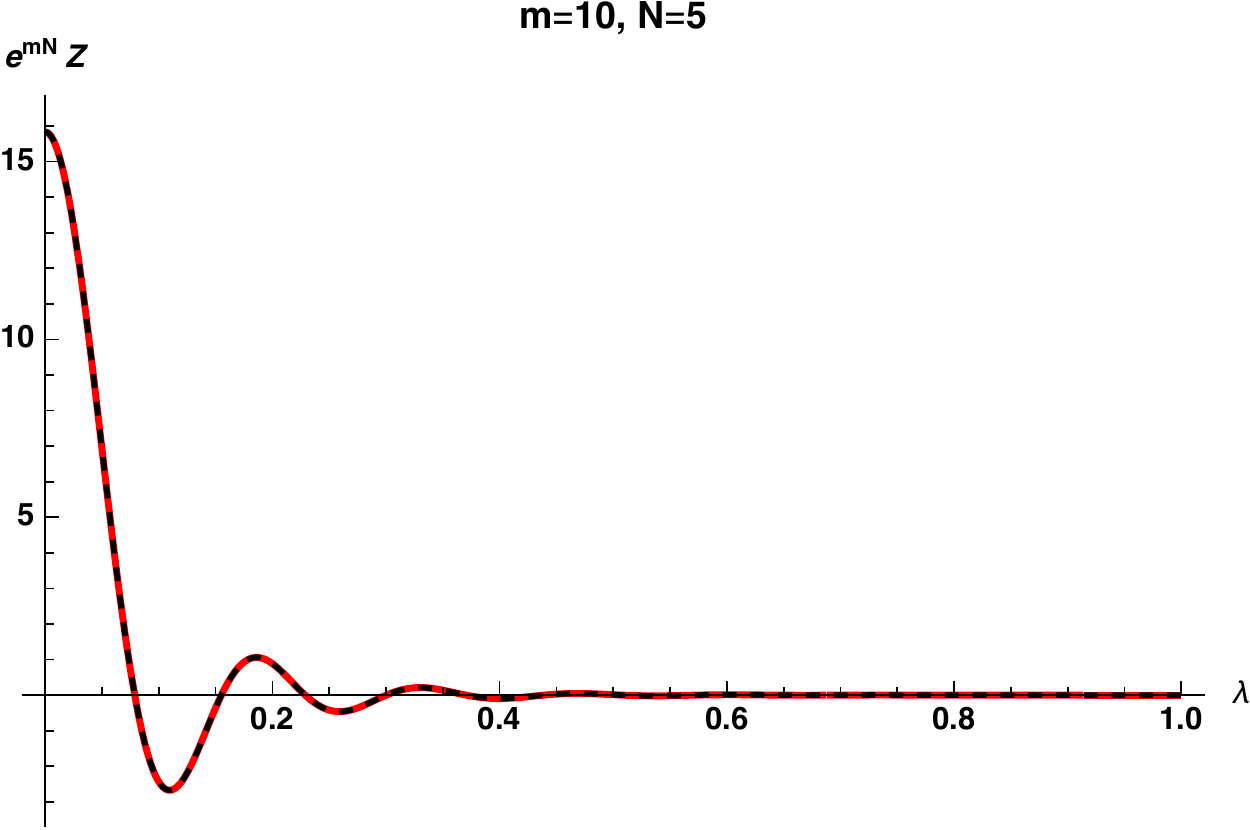} 
\caption{Approximation of $e^{mN} \mathcal{Z}_N ^{U(1)}$ at large $m=10$ as
a function of $\protect\lambda= \protect\eta/N$, using \eqref{eq:largem1}
(red) and \eqref{eq:largem2} (black, dashed), for $N=2$ (above) and $N=5$
(below).}
\label{fig:largemass}
\end{figure}

We can represent the function \eqref{eq:exactz1} in yet another form, in
terms of a conical function \cite{GilSegura,Russo:2016ueu}: 
\begin{equation*}
\mathcal{Z}_N ^{U(1)}= \frac{\sqrt{2 \pi} }{2^N (\sinh(m))^{N-\frac{1}{2}} } 
\frac{ \Gamma (N + i \eta ) \Gamma (N - i \eta ) }{ \Gamma (N )} P^{\frac{1}{%
2} -N } _{- \frac{1}{2} + i \eta} (z) ,
\end{equation*}
where $P^{\frac{1}{2} -N } _{- \frac{1}{2} + i \eta} (z) $ is an associated
Legendre function of negative order and complex degree. This latter form is
the most suitable to study the asymptotics for large mass. Indeed, when $m
\to \infty$, $z=\cosh (m) \to \infty$ as well and we can use the
approximation of \cite{Dunster:1991}: 
\begin{align*}
P^{\frac{1}{2} -N } _{- \frac{1}{2} + i \eta} (z) & \approx \sqrt{ \frac{2
\pi}{z}} \frac{ \sin \left( \eta \log (2 z) + \theta_1 + \theta_2 \right) }{
\sinh (\pi \eta) \lvert \Gamma (1 + i \eta) \Gamma (N + i \eta) \rvert } 
\notag \\
& = \sqrt{ \frac{2 }{\pi z}} \frac{ \sin \left( \eta \log (2 z) + \theta_1 +
\theta_2 \right) }{ \prod_{k=0} ^{N-1} \sqrt{ k^2 + \eta^2 } } ,
\end{align*}
where $\theta_1 = \arg \Gamma (1 + i \eta)$ and $\theta_2 = \arg \Gamma (N -
i \eta)$, and in the second line we used elementary identities for the $%
\Gamma$ function. Altogether, and approximating the hyperbolic functions for 
$m \to \infty$, we have: 
\begin{equation}  \label{eq:largem1}
\mathcal{Z}_N ^{U(1)} \approx \frac{ e^{-m N} \pi \prod_{k=1}^{N-1} \sqrt{%
k^2 + \eta^2} }{2^{N-1} \Gamma (N) \sinh (\pi \eta) } \sin \left( \eta m +
\theta_1 + \theta_2 \right) .
\end{equation}
This approximation is in agreement with the large mass approximation found
in \cite{Russo:2016ueu} (Eq. (8) therein) applying a different Euler
transformation to \eqref{eq:exactz1}, which led to: 
\begin{equation}  \label{eq:largem2}
\mathcal{Z}_N ^{U(1)} \approx \frac{2 \pi e^{-m N} }{ \Gamma (N)
\sinh (\pi \eta) } \Im \left( e^{i m \eta} \prod_{k=1} ^{N-1} (k-i\eta)
\right) .
\end{equation}
See figure \ref{fig:largemass} for the match of expressions \eqref{eq:largem1}
and \eqref{eq:largem2}.

The exact evaluation \eqref{eq:exactz1} of the partition function, or its equivalent representation as a conical function, relies on the hypothesis $\cosh (m) \ge 1$, thus on reality of the mass. However, the dependence of $\mz_N ^{U(1)} $ on $m$ should be holomorphic \cite{Jafferis:2010un,Closset:2012vg}. For arbitrary complex masses the integral \eqref{eq:z1} can be evaluated by residue theorem \cite{Benvenuti:2011ga}, and we checked for many values of $N$ that the result coincide with the prolongation of \eqref{eq:exactz1} to complex masses.

\paragraph*{Integrability.}

The partition function satisfies the second-order differential equation \cite{Russo:2016ueu}
\begin{equation}
\label{eq:diffeq}
    \frac{d^2 \mz_N}{dm^2} + 2N \coth (m) \frac{d \mz_N}{dm } + (\eta^2 + N^2) \mz_N = 0,
\end{equation}
which becomes the Schr\"{o}dinger equation with a hyperbolic P\"{o}schl-Teller potential, for the function $Z(m) = \sinh(m)^N \mz_N$ \cite{Russo:2016ueu}. This quantum mechanical model has a discrete energy spectrum \cite{Flugge}, and $Z(m)$ represents the wave function of a state with positive energy proportional to $\eta^2$. Furthermore, the fact that the potential appears with integer coefficient $N$ implies that the wave function propagates without reflection.

The appearance of the quantum mechanical interpretation with a solvable P\"{o}schl-Teller potential immediately suggests a possible role of the hyperbolic Calogero-Moser model, the celebrated integrable system, which can
be seen as the many-body generalization of the quantum mechanical problem
above. The Hamiltonian of the $A_{\widehat{N}-1}$ hyperbolic
Calogero-Moser model is \cite{Olshanetsky:1981dk,HR2} 
\begin{equation}
H=\sum_{1 \le j<k \le \widehat{N}}\left[ -\hbar^{2} \partial
_{x_{j}} \partial_{x_{k}} +\frac{g(g-\hbar) \mu ^{2}}{4\sinh
^{2}\left( \mu \left( x_{j}-x_{k}\right) /2\right) } \right],  \label{H}
\end{equation}%
and there exists $\widehat{N}-1$ additional independent partial
differential operators $H_{l}$ of order $l$, such that the PDOs form a
commutative family. The simplest is the momentum operator%
\begin{equation}
H_{1}=-i \hbar \sum_{j=1}^{\widehat{N}}\partial _{x_{j}}  \label{P}
\end{equation}%
whereas the others are made of correspondingly higher derivatives (and lower
order terms as well). Here, $\widehat{N} = n+1$. Consider the two-particle case, the family is then the
Hamiltonian and the momentum operator, \eqref{H} and \eqref{P}.

The result in what follows appears to have some similitudes with the work \cite%
{Isachenkov:2016gim} (further extended in \cite%
{Isachenkov:2017qgn,Isachenkov:2018pef}) where conformal blocks of scalar
4-point functions in $d$-dimensional conformal field theory are mapped to
eigenfunctions of the two particle hyperbolic Calogero-Moser
system. The relevant model there corresponds to the $BC_{2}$ case rather
than the $A_{1}$ or $A_{\widehat{N}-1}$ here (see below), due to the
orthogonal symmetry there.

Using recent work on the construction, by a recursive method, of the joint
eigenfunctions of this integrable system \cite{HR2}, we show now that the
Abelian theory above can be identified with this two-particle $A_{1}$
hyperbolic Calogero-Moser, where the coupling constant $g$ in \eqref{H} will be identified with the
half-number of flavours $N$. In particular, this two-particle interpretation follows
from considering the function%
\begin{equation*}
\Psi _{2}(g;x,y)\equiv e^{iy_{2}(x_{1}+x_{2})}\int_{-\infty }^{\infty
}e^{i(y_{1}-y_{2})z}K_{2}(g;x,z)dz,
\end{equation*}%
where the kernel, with $g>0$, $x,y\in \mathbb{R}^{2}$, is 
\begin{equation*}
K_{2}(g;x,z)=\frac{\left[ 4\sinh ^{2}\left( x_{1}-x_{2}\right) \right] ^{g/2}%
}{\prod\limits_{j=1}^{2}\left[ 2\cosh \left( x_{j}-z\right) \right] ^{g}},
\end{equation*}%
and is central in the recursion, taking the $\widehat{N}-1$ eigenfunction
to the $\widehat{N}$ eigenfunction. The connection with the function $Z(m)$ defined above follows immediately from the identifications $g=N$, $x_1 = m/2 = -x_2$ and $(y_1-y_2)/2 = \eta$.
It is shown in \cite{HR2} that 
\begin{align*}
H_{1}\Psi _{2}(x,y)& =(y_{1}+y_{2})\Psi _{2}(x,y), \\
H\Psi _{2}(x,y)& =(y_{1}^{2}+y_{2}^{2})\Psi _{2}(x,y).
\end{align*}%

A different type of connection also exists relating the non-Abelian theory, with $%
\widehat{N}=N$, with the free case of the integrable system, given by $g=\hbar$ in \eqref{H}. Using the customary adimensional coupling $%
\widehat{\lambda }\equiv g/ \hbar=1$, \eqref{H} is then the free $N$%
-body Hamiltonian. Thus, there is no identification here between $g$ and
number of flavours and is a very different relationship compared to the
two-particle one. The integral representation given for $\Psi _{N}(\widehat{\lambda} ;x,y)$ \cite{HR2} is then evaluated exactly for $\widehat{%
\lambda }=1$ and the explicit expression \cite[Theorem 3.1.]{HR2} is the one
for the partition function of the $T[SU(N)]$ linear quiver \cite{Nishioka:2011dq,Benvenuti:2011ga,Gulotta:2011si}.

The relationship between the integral expressions in \cite{HR2} and the well-known Heckman-Opdam hypergeometric functions \cite{HO}, which are also relevant in \cite{Isachenkov:2016gim,Isachenkov:2017qgn}, is explained in \cite{HR2}. By factorizing $\Psi _{N}$ in two pieces, one describing the centre of mass, it is shown in \cite{HR2} that the remaining piece is
the $A_{N-1}$ Heckman-Opdam hypergeometric function. In terms of two sets of $N$ variables $\left(m_{j}, \zeta_j \right)_{j=1} ^N$, this hypergeometric satisfies the condition $\sum_{j} m_j =0 = \sum_{j} \zeta _j$, with $\zeta_j \in \R$ and complex $m_j$ such that $\lvert \Im (m_j - m_k) \rvert < \pi$, cfr. \cite[Theorem 7.1]{HR2}. On the gauge theory side, those are exactly the constraints on the $T[SU(N)]$ theory \cite{Benvenuti:2011ga}, the first being the $SU(N)$ flavour symmetry and the latter arising from the redundancy of the $N$ number of $\zeta_j$ variables, defined from the original $N-1$ FI parameters as $\zeta_j = \eta_j - \eta_{j+1}$ \footnote{Equivalently, in terms of D-brane construction, $(m_j, \zeta_j)$ are on equal footage, seen as backgrounds living on two separated regions of a D3-brane \cite{Gulotta:2011si}, thus the $SU(N)$ constraint should be imposed on both.}. We underline that the partition function of the $T[SU(N)]$ quiver is evaluated for real masses and FI parameters, but can, by holomorphicity, hold on the stripes $\lvert \Im (m_j - m_k) \rvert < \pi$, hence the identification is exact.

\paragraph*{Abelian theory at large $N$.}

Sending $N \to \infty$ in the double scaling limit with $\lambda \equiv \eta/N$ fixed, the leading contribution to the partition function %
\eqref{eq:z1} comes from the saddle points of the action 
\begin{equation}  \label{eq:s1x}
S_1 (x) = - i \lambda x + \frac{ \sinh (x) }{ \cosh (x) + z } ,
\end{equation}
which are given by the set $\mathscr{S} = \left\{ x_s ^{\pm} + i 2 \pi k , \
k \in \mathbb{Z} \right\}$, with 
\begin{equation}  \label{eq:xs1}
x_s ^{\pm} = \log \left( \frac{ - \lambda z \pm i \Delta }{ i + \lambda }
\right) ,
\end{equation}
where $\Delta \equiv \sqrt{1- \lambda^2 \sinh (m) ^2 }$ and we recall that $%
z \equiv \cosh (m)$. The curve $\lambda \sinh (m) =1$ determines a critical
line in parameter space, along which the free energy $\mathcal{F} = - \frac{1%
}{N} \log \mathcal{Z}$ has a discontinuity in its second derivative. In the 
\emph{sub-critical phase} $\lambda \sinh(m)<1$, the leading contribution
comes from $x_s^{+}$ and $k=0$, while in the \emph{super-critical phase} $%
\lambda \sinh(m) >1$ both $x_s ^{\pm}$ contribute, being complex conjugate
and $S_1 (x_s ^{-}) = S_1 (x_s ^{+} )^{\ast}$.

Close to the saddle points $\bar{x} \in \mathscr{S}$, we can change
variables $x= \bar{x} + t / \sqrt{N}$ and expand 
\begin{equation*}
S_1 (x) = S_1 ( \bar{x} ) + \frac{t^2 S^{\prime \prime } _{1} (\bar{x}) }{2 N%
} + \frac{ t^3 S^{\prime \prime \prime } _1 (\bar{x} ) }{6 N^{\frac{3}{2} } }
+ \frac{t^4 S^{ (iv) } _{1} (\bar{x}) }{24 N^2} + \dots
\end{equation*}
We now plug this expansion into \eqref{eq:z1} and keep the Gaussian part in $%
t$ exponentiated, while expanding the rest of the exponential function.
Elementary integration provides: 
\begin{align*}
\mathcal{Z} ^{U(1)} & = 2^{-N} \sqrt{ \frac{ 2 \pi }{N}} \sum_{\bar{x} \in %
\mathscr{S} } \frac{ e^{- N S_1 ( \bar{x} ) } }{ \sqrt{ S^{\prime \prime }
_1 (\bar{x} ) } } \left[ 1 + \right. \\
& \left. \frac{1}{24 N} \left( \frac{ 5 S^{\prime \prime \prime } _1 (\bar{x}%
)}{ (S^{\prime \prime } _1 (\bar{x}))^3 } - \frac{ 3 S^{ (iv) } _1 (\bar{x})%
}{ (S^{\prime \prime } _1 (\bar{x}))^2 } \right) + \mathcal{O} (N^{-2}) %
\right] .
\end{align*}
The relevant expressions for the derivatives of the action $S_1$ are
reported in the Appendix \hyperlink{app:derivatives}{A}. When $\lambda \sinh
(m) <1$, only $x_s ^{+}$ contributes, and we get: 
\begin{align*}
\mathcal{Z}^{U(1)}_ {\mathrm{sub.}} & = 2^{-N} \sqrt{ \frac{ 2 \pi }{N}} 
\frac{ e^{- N S_1 ( x_s ^{+} ) } }{ \sqrt{ S^{\prime \prime } _1 (x_s ^{+} ) 
} } \left[ 1 + \right. \\
& \left. \frac{1}{24 N} \left( \frac{ 5 S_1 ^{\prime \prime \prime } (x_s^+
) }{ (S_1 ^{\prime \prime} (x_s^+ ))^{3}} - \frac{ 3 S_1 ^{(iv)} (x_s^+ ) }{
(S_1 ^{\prime \prime} (x_s^+ ))^{2}} \right) \right] + \mathcal{O} (N^{-2}) ,
\end{align*}
while in the supercritical phase $\lambda \sinh (m) >1$ both $x_s ^{\pm}$
must be taken into account, leading to: 
\begin{equation*}
\mathcal{Z}^{U(1)}_ {\mathrm{super.}} = 2 \Re \left( \mathcal{Z}^{U(1)}_ {%
\mathrm{sub.}} \right) + \mathcal{O} (N^{-2}) .
\end{equation*}
Dropping sub-leading corrections, one can evaluate $\mathcal{F}$ in both
phases: 
\begin{equation}  \label{eq:F1}
\mathcal{F}^{U(1)}_ {\mathrm{sub.}} = S_1 (x_s ^{+}), \quad \mathcal{F}%
^{U(1)}_ {\mathrm{super.}} = \Re \left( S_1 (x_s ^{+}) \right) ,
\end{equation}
with discontinuous second derivative: 
\begin{equation*}
\frac{ \partial ^2 \mathcal{F}^{U(1)}_ {\mathrm{sub.}} }{ \partial \lambda^2}
- \frac{ \partial ^2 \mathcal{F}^{U(1)}_ {\mathrm{super.}} }{ \partial
\lambda^2} = \frac{z}{(1+\lambda^2) \Delta} .
\end{equation*}
Therefore, not only the susceptibility $\frac{ \partial ^2 \mathcal{F} }{%
\partial \lambda^2} $ is discontinuous, but it is divergent as $%
(\lambda-\lambda_c)^{-\gamma_c}$, and we identify the critical exponent $%
\gamma_c = \frac{1}{2}$. The free energy yields analogous discontinuity with
respect to the mass: 
\begin{equation*}
\frac{ \partial ^2 \mathcal{F}^{U(1)}_ {\mathrm{sub.}} }{ \partial m^2} - 
\frac{ \partial ^2 \mathcal{F}^{U(1)}_ {\mathrm{super.}} }{ \partial m^2} = 
\frac{ z \Delta}{\sinh (m)^2} - \frac{\lambda z}{ \Delta} ,
\end{equation*}
hence the critical exponent for the mass is again $\delta_c=\frac{1}{2}$.

\begin{figure}[bth]
\centering
\includegraphics[width=0.4\textwidth]{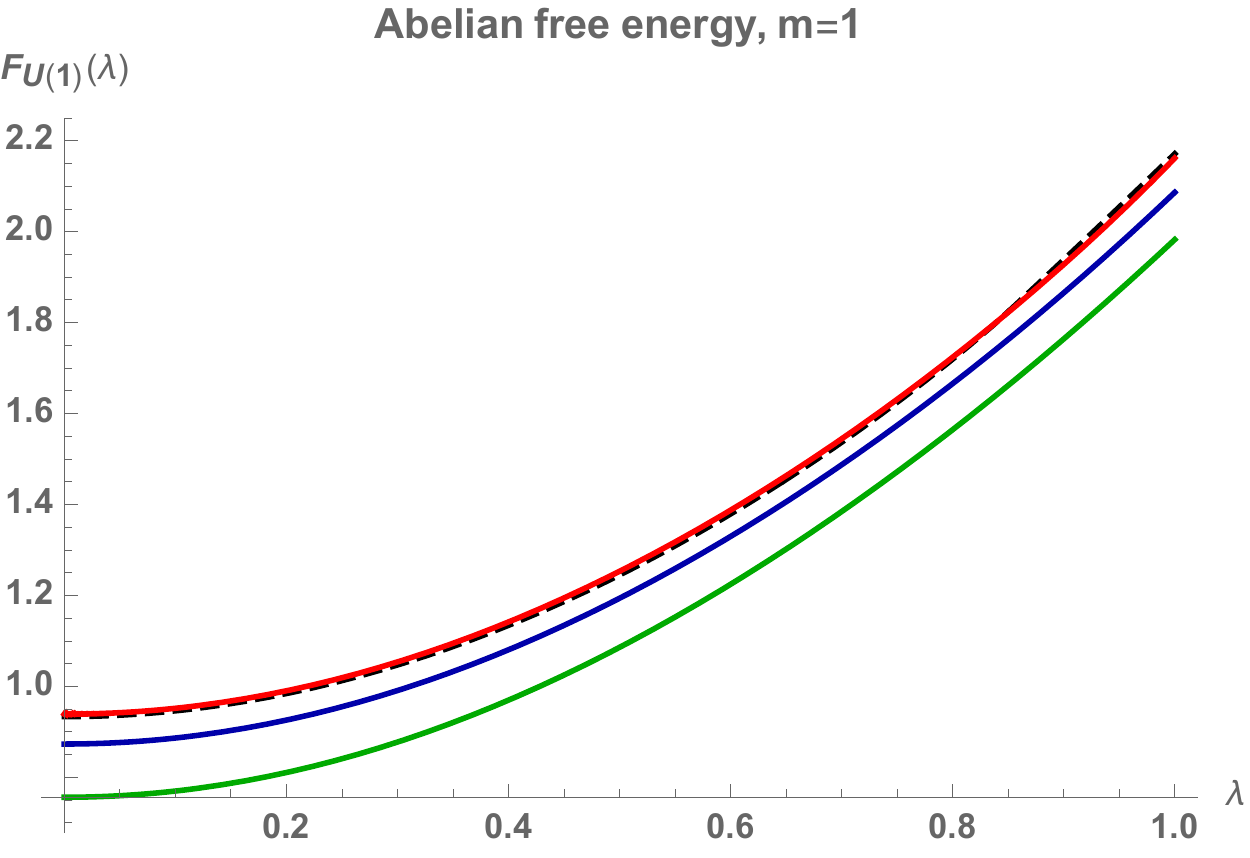} 
\caption{Exact solution of $\mathcal{F}^{U(1)}$ as a function of $\protect%
\lambda = \protect\eta/N$ at $m=1$, for $N=4,7,20$ (in green, blue, red,
respectively) and large $N$ expression (black, dashed).}
\label{fig:abelianF}
\end{figure}

In figure \ref{fig:abelianF} we present the convergence of the exact
solution \eqref{eq:exactz1} and the large $N$ expression \eqref{eq:F1} as $N$
is increased.

\paragraph*{Wilson loops.}

Irreducible complex representations of $U(1)$ are labelled by $r \in \mathbb{%
Z}$, thus Wilson loops can be written as $W_r= \mathrm{Tr}_r e^{x} = e^{rx}$
(recall that the radius of the three-sphere is $1/2\pi$), and their
expectation value is: 
\begin{align}  \label{eq:WL1}
\langle W_r \rangle & = \frac{1}{2^N \mathcal{Z}_N ^{U(1)} } \int_{- \infty}
^{+ \infty} dx \frac{e^{(i \eta+r) x } }{ \left[ \cosh (x) + z \right]^{N} }
\notag \\
& = \frac{ \Gamma (N +r + i \eta ) \Gamma (N -r - i \eta ) }{ \Gamma (N+
i\eta ) \Gamma \left( N - i \eta \right) }  \notag \\
& \times \frac{ {_2 F _1 } \left( \frac{1}{2} -r - i \eta, \frac{1}{2} +r +
i \eta , N + \frac{1}{2} , \frac{1-z}{2} \right) }{ {_2 F _1 } \left( \frac{1%
}{2} - i \eta, \frac{1}{2} + i \eta , N + \frac{1}{2} , \frac{1-z}{2}
\right) },
\end{align}
where we stress that the insertion of a Wilson loop is analogous to the
complexification of the FI coupling. The integral
representation \eqref{eq:WL1} is well-defined as $\eta \to 0$ only for
representations of size $\lvert r \rvert < N$: this is reflected in the
poles of the $\Gamma$ function at negative integers.

The quantum mechanical interpretation carries over for the
Wilson loop without FI term, $\eta =0$. In this case, $w_r \equiv \left[
\sinh (m)^N \mathcal{Z}_N \langle W_r \rangle\right]_{\eta=0}$ satisfies the Schr\"{o}dinger equation with P\"{o}schl-Teller potential: 
\begin{equation*}
\left[  \frac{ d^2 \ }{ dm^2} - \frac{N (N-1)}{\sinh(m)^2} \right] w_r  = r^2 w_r .
\end{equation*}
The latter equation describes the wave function of a bound state with energy proportional to $r^2$, for integer $\lvert r \rvert < N$, which is indeed the case at hand \cite{Flugge}.

For $\eta \ne 0$, however, the resulting potential acquires an imaginary
part, seemingly spoiling unitarity of the evolution operator and producing a
dissipation-like term in the probability conservation.

At large $N$ with the size $r$ of the representation fixed, the Wilson loop
can be approximated by the value of the integrand in \eqref{eq:WL1} at the
saddle points. Nevertheless, we can also consider the case of large
representations, in which $r$ scales with $N$, \textit{i.e.} $f\equiv r/N $
is kept fixed as $N \to \infty$. Let us turn off the FI term for simplicity, 
$\eta=0$, the saddle points of the action are given by: 
\begin{equation*}
\bar{x} = \log \left( \frac{ f \cosh (m) \pm \sqrt{ 1 + f^2 \sinh(m) ^2 } }{
1- f} \right) + i 2 \pi k ,
\end{equation*}
with $k \in \mathbb{Z}$, that are real for every $-1 < f < 1$\footnote{%
As $\lvert r \rvert < N$, the range of validity is $-1 < f < 1$. $x_s ^{+}$
is singular as $f \to 1^{-}$, while $x_s^{-}$ is singular as $f \to -1^{+}$.}%
. Therefore, the Wilson loops without FI term do not experience phase
transition. The limit with both $\eta$ and $r$ scaling with 
$N$ is commented in Appendix \hyperlink{app:WLscaling}{B}.

\paragraph*{$J_3$ correlators.}
We can also consider other families of operators, besides Wilson loops. Higgs branch operators in $3d$ $\mn =4$ can be analyzed through localization techniques \cite{Dedushenko:2016jxl}, and therefore represent a suitable choice for the present setting.
In particular, we focus our attention on the gauge invariant, quadratic operator
\begin{equation*}
    J_3 = \frac{1}{N} \left[ \widetilde{Q}_{+,j} Q_{+} ^{j} -  \widetilde{Q}_{-,j} Q_{-} ^{j}  \right] ,
\end{equation*}
where $Q_{\pm,j}$, $j=1, \dots, N$, are the hypermultiplets of mass $\pm m$. The expectation value of this operator is \cite{Russo:2016ueu}
\begin{equation*}
    \langle J_3 \rangle = \frac{1}{2N \mz_N} \frac{ d \mz_N }{ dm} ,
\end{equation*}
and correlation functions of $J_3$ are generated by higher derivatives.

The differential equation \eqref{eq:diffeq} satisfied by $\mz_N$ can be translated into a recursion relation for correlators of $J_3$:
\begin{equation*}
    \langle J_3 J_3 \rangle = - \coth (m) \langle J_3 \rangle - \frac{1}{4N} \left(1+ \frac{ \eta^2 }{ N^2} \right) .
\end{equation*}
Taking the first derivative of Eq. \eqref{eq:diffeq} gives $\frac{ d^3 \mz_N }{dm^3}$ as a function of the first and second derivative of $\mz_N$, but the second order term can be eliminated using \eqref{eq:diffeq}. Hence, we immediately obtain:
\begin{align*}
    \langle J_3 J_3 J_3 \rangle & = \langle J_3 \rangle \left[ \frac{2N \cosh(m)^2 + 1}{2N \sinh (m)^2 } - \frac{1}{4} \left( 1 + \frac{ \eta^2}{N^2} \right) \right] \\
    & +  \frac{1}{4} \left( 1 + \frac{ \eta^2}{N^2} \right) .
\end{align*}

One can take further derivatives and systematically plug \eqref{eq:diffeq} in the resulting expression. This allows to recursively compute $k$-point correlation functions of $J_3$: exploiting Eq. \eqref{eq:diffeq}, the final result will be an expression only in terms of $\langle J_3 \rangle$, hyperbolic functions of $m$ and polynomials in $\left( 1+ \eta^2 /N^2 \right)$.

\paragraph*{Non-Abelian theory: $SU(2)$.}

The simplest non-Abelian theory corresponds to the gauge group $SU(2)$. The
partition function is again a single integral, but now the one-loop
determinant of the vector multiplet contributes. Also, the $SU(2)$ vector
multiplet cannot be coupled to an FI background, therefore $\eta=0$%
. The partition function is: 
\begin{equation*}
\mathcal{Z}_N ^{SU(2)} = \int_{- \infty} ^{+ \infty} dx \frac{ \sinh (x) ^2 
}{2^N \left[ \cosh (x) + z \right]^{N} } .
\end{equation*}
Writing $\sinh (x)$ in terms of exponentials, we can see the $SU(2)$
partition function as a combination of expectation values of Wilson loops in
the Abelian theory: 
\begin{equation*}
\mathcal{Z}_N ^{SU(2)} = \left[ \frac{\mathcal{Z}_N ^{U(1)}}{2} \left(
\langle W_2 \rangle -2 + \langle W_{-2} \rangle \right) \right]_{\eta=0} ,
\end{equation*}
with the expectation value $\langle W_r \rangle$ given in Eq. \eqref{eq:WL1}.

Due to the absence of FI term, the unique saddle point is $x_{s}=0$, and the phase structure at large $N$ is trivial.

\paragraph*{Non-Abelian theory: $U(2)$.}

We now apply the same procedure to the $U(2)$ theory, \textit{i.e.} two
colours. Specialization of \eqref{eq:znc} for $n=2$ gives: 
\begin{equation}  \label{eq:z2}
\mathcal{Z} ^{U(2)} _N = \int_{\mathbb{R}^2} \frac{ e^{i \eta (x_1 + x_2) }
\left( 2 \sinh \frac{x_1 - x_2}{2} \right)^2 dx_1 dx_2 }{ 2^{2N} \left[
\left( \cosh (x_1) + z \right)\left( \cosh (x_2) + z \right) \right]^N },
\end{equation}
where, as above, $z \equiv \cosh (m)$. Through the equivalent representation
of \eqref{eq:z2} as a determinant, one could write an exact solution 
\begin{equation*}
\mathcal{Z} ^{U(2)} _N = 2! \det_{1 \le j,k \le 2} [ Z_{jk} ],
\end{equation*}
with $Z_{jk}$ entries of a $2 \times 2$ matrix formally given by %
\eqref{eq:exactz1} up to a shift in the FI coupling $i \eta \mapsto i \eta +
j + k -2$, $j,k \in \left\{ 1,2 \right\}$. This equals the
determinant of a matrix whose entry $(j,k)$ is the expectation value, in the
Abelian matrix model, of a Wilson loop in the irreducible representation
labelled by $j+k-2$: 
\begin{equation*}
\mathcal{Z} ^{U(2)} _N = 2 (\mathcal{Z}_N ^{U(2)})^2 \left( \langle W_2 \rangle
- \langle W_1 \rangle^2 \right) .
\end{equation*}

To study \eqref{eq:z2} in the limit in which the number of flavours $N$ is
large, we notice that the interaction between eigenvalues is sub-leading in $%
1/N$, thus the saddle points of the $U(2)$ theory are those of the action $%
S_1 (x_1) + S_1 (x_2)$: 
\begin{equation*}
\mathscr{S}^2 = \left\{ (x_s ^{\pm} + 2 \pi k_1, x_s ^{\pm} + 2 \pi k_2), \
k_{1,2} \in \mathbb{Z} \right\} .
\end{equation*}
We proceed as in the Abelian case: we change variables $x_{1,2} = \bar{x}%
_{1,2} + t_{1,2} /\sqrt{N}$ and expand both the action and the hyperbolic
interaction around the saddle point $(\bar{x} _1, \bar{x} _2)$. Expanding up
to $\mathcal{O} (N^{-1})$ and integrating we obtain, for the \emph{%
sub-critical phase}: 
\begin{align*}
\mathcal{Z} ^{U(2)} _{\mathrm{sub.}} & = \frac{\pi}{2^{2^(N-1)} N^2 } \frac{
e^{-2N S_1 (x_s ^{+})} }{ (S^{\prime \prime} _1 (x_s ^{+}))^2 } \left[ 1 +
\right.  \notag \\
& + \left. \frac{1}{2 N} \left( \frac{1}{ S^{\prime \prime} _1 (x_s ^{+}) }
+ \frac{ 17 \left( S^{\prime \prime \prime } _1 (x_s^{+}) \right)^2 }{6
(S^{\prime \prime} _1 (x_s ^{+}))^3 } - \frac{ 3 S^{(iv) } _1 (x_s^{+}) }{2
(S^{\prime \prime} _1 (x_s ^{+}))^2 } \right) \right] ,
\end{align*}
while the expression in the \emph{super-critical phase} $\lambda \sinh(m)>1$
is a sum of four pieces, and is reported in Appendix \hyperlink{app:Z2super}{%
C}.

Dropping $1/N$ corrections, the free energy is simply $\mathcal{F} ^{U(2)} =
2 \mathcal{F} ^{U(1)}$, in particular the phase transition is second order
with the same critical exponent $\gamma_c=\frac{1}{2}$. In figure \ref%
{fig:NAbF} we show how the exact solution approaches the large $N$
expression as $N$ is increased.

We study the most general non-Abelian case in Appendix \hyperlink{app:Unc}{D}%
, and only report here the main result. The free energy at large $N$ of the $%
U(n)$ theory is $n$ times the free energy of the Abelian theory: 
\begin{equation*}
\mathcal{F} ^{U(n)} = n \mathcal{F} ^{U(1)} .
\end{equation*}

\begin{figure}[bth]
\centering
\includegraphics[width=0.4\textwidth]{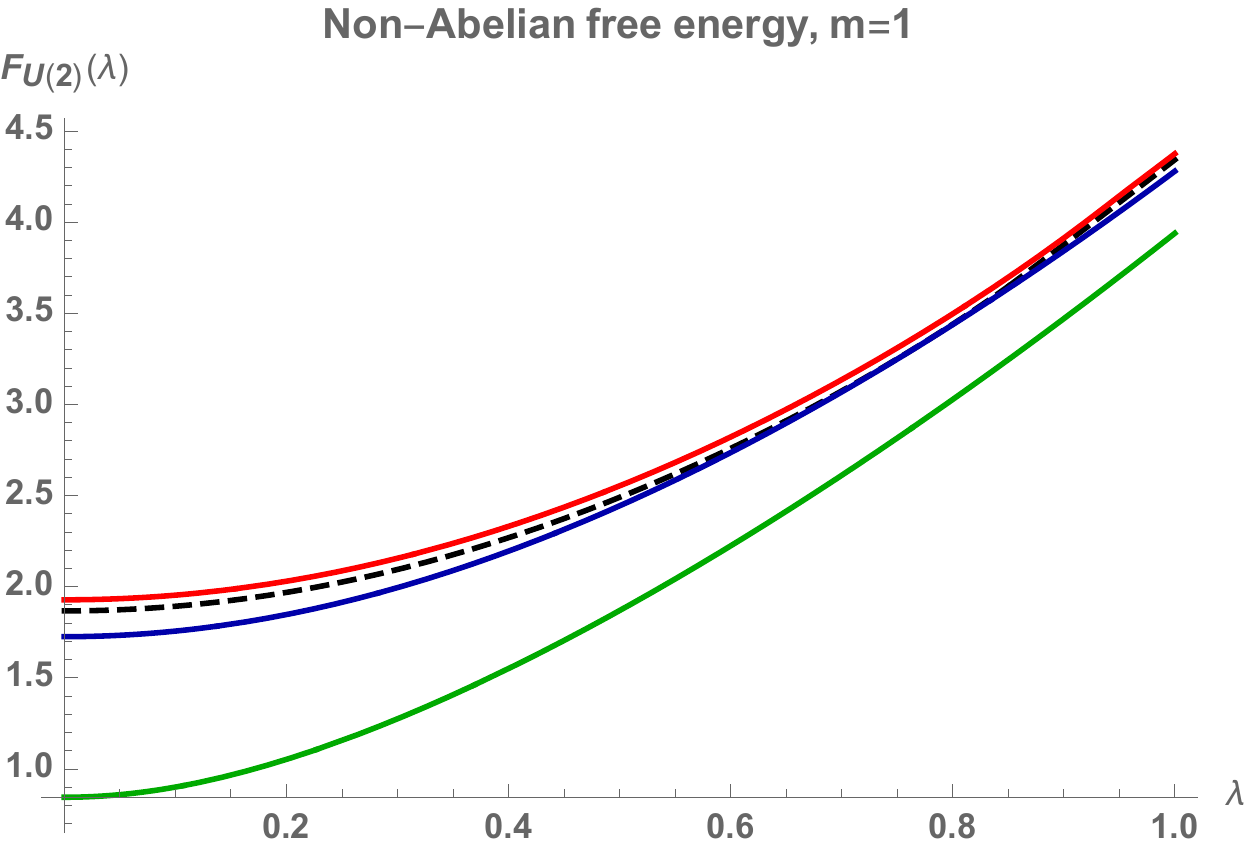} 
\caption{Exact solution from determinants of $\mathcal{F}^{U(2)}$ as a
function of $\protect\lambda = \protect\eta/N$ at $m=1$, for $N=4,7,100$ (in
green, blue, red, respectively) and large $N$ expression (black, dashed).}
\label{fig:NAbF}
\end{figure}

\paragraph*{Other $R$-charges.}

To conclude, we show how the features of the $\mathcal{N} =4$ theory with $2N
$ chiral multiplets with $R$-charge $q= \frac{1}{2}$ can be extended to the $%
\mathcal{N} =2 $ theory with $2N$ chiral multiplets with more general
assignment of $R$-charge $q$. The expressions for the partition function and
the saddle point equation for arbitrary $q$ are reported in Appendix 
\hyperlink{app:Rcharge}{E}. Here we comment on how the theory at
half-integer $q \in \frac{1}{2} \mathbb{Z}$ can be obtained by simple
modification of the results in \cite{Russo:2016ueu}.

$q=1$. In this case the action is pure imaginary, already at finite $N$, and
admits no saddle point.

$q \in \frac{1}{2} + \mathbb{Z}$. The saddle point equation reduces to:  
\begin{equation*}
\frac{ \sinh (x) }{ \cosh (x) + z } = \frac{ i \lambda}{2 (1-q)} ,
\end{equation*}
and the large $N$ behaviour is identical to the case $q= \frac{1}{2}$ upon
scaling $\lambda \mapsto \frac{ \lambda}{2 ( 1-q)}$.

$q \in \mathbb{Z} \setminus \left\{ 1 \right\}$. For integer non-unit $q$
the saddle point equation simplifies into:  
\begin{equation*}
\frac{ \sinh (x) }{ \cosh (x) - z } = \frac{ i \lambda}{2 (1-q)} ,
\end{equation*}
and the phase structure at large $N$ is identical to the case $q = \frac{1}{2%
}$, up to scaling $\lambda \mapsto \frac{ \lambda}{2 ( 1-q)}$ and replace in
the formulae $z \mapsto -z$. The critical line is $\lambda \sinh (m) = 2
\lvert 1-q \rvert$.

As a future direction, it would be interesting to study the large $N$ free energy for more general $R$-charges and determine the $R$-symmetry in the IR by $\mf$-extremization \cite{Jafferis:2010un,Closset:2012vg}. A crucial question then would be whether there exists more than one solution $q_{\mathrm{IR}}$, and analyze the corresponding theories as a function of $\lambda$, along the lines of \cite{Gukov:2015qea,Gukov:2016tnp}.

\paragraph*{Acknowledgements.}
    We thank Luis Melgar and Jorge Russo for discussions and correspondence. The work of MT was supported by the Funda\c{c}\~{a}o para a Ci\^{e}ncia e a Tecnologia (FCT) through IF/01767/2014. The work of LS was supported by the FCT through SFRH/BD/129405/2017. The work is also supported by FCT Project PTDC/MAT-PUR/30234/2017.

\bibliography{Phases_QED3.bib}

\onecolumngrid

\begin{appendix}

\hypertarget{app:derivatives}{\section{Appendix A. Derivatives of the action $S_1 $}}

Here we present the full expressions for the derivatives of the action in the Abelian theory, evaluated at the saddle point $\bar{x}= x_s^+$. In what follows, we denote $z \equiv \cosh (m)$, $\xi \equiv \lambda \sinh (m)$ and $\Delta \equiv \sqrt{1-\lambda^2 \sinh (m)^2}$.
\begin{align*}
	S_1 (x_s ^+) & = \log \left( \frac{\Delta z + i \lambda \sinh (m)^2 + 1 }{ \left( i + \lambda \right) \left(  \lambda z - i  \Delta  \right) } \right) - i \lambda \log \left( \frac{ - \lambda z + i \Delta  }{i + \lambda } \right) , \\
	S_1 ^{\prime \prime} (x_s ^+) & = \lambda^2 \left[ 1 + \frac{ z \Delta  -1 }{ \xi^2 } \right] , \\
	S_1 ^{\prime \prime \prime } (x_s ^+) & = \frac{\lambda  \left( 1-i \lambda \right)}{4\sinh (m)^2 \left( \lambda \sinh (m)^2  - i z \Delta  - i \right) ^{2}} \left[ 2\xi -6 \xi z \Delta +2\left( 4\xi ^{2}-3\right) \xi \cosh (2m) +8 i \xi ^{2}\sinh ( m)^3  \right.\notag \\
			& \left.  +2\xi \cosh (3m) \Delta   + 7 i \sinh (m) - 8i \xi ^{2} \sinh (2m) \Delta   +2i \sinh (2m) \Delta - i \sinh (3m) \right] , \\
	S_1 ^{(iv)} (x_s ^+) & = - 6 \lambda^4 + \frac{23  \lambda^2  }{2 \sinh^2 m } - \frac{ \Delta \left( 48 \lambda^2 \sinh (m)^2 - 23 \right) z  + 2 \left( 7 \lambda^2 \sinh (m)^2 -4 \right) \cosh ( 2m)  + \Delta \cosh (3m)  - 16 }{ 4 \sinh^4 m } . 
\end{align*}
The values of the derivatives of $S_1$ when evaluated at $x_s ^{-}$ are immediately obtained through the relations:
\begin{equation*}
		S_1 (x_s^-) = (S_1 (x_s^+))^{\ast} , \quad S_1 ^{\prime \prime} (x_s^-) = (S_1 ^{\prime \prime} (x_s^+))^{\ast} , \quad S_1 ^{\prime \prime \prime } (x_s^-) = - (S_1 ^{\prime \prime \prime} (x_s^+))^{\ast} , \quad S_1 ^{(iv)} (x_s^-) = (S_1 ^{(iv)} (x_s^+))^{\ast}
\end{equation*}

\hypertarget{app:WLscaling}{\section{Appendix B. Multiple scaling limit of Wilson loops}}
The large $N$ limit of Eq. \eqref{eq:WL1} whit $\lambda \equiv \eta/N$ and $f \equiv r/N$ fixed, with $0\le f <1$, is obtained from the contributions of the saddle points:
\begin{equation*}
	\bx = \log \left( \frac{ - \left(  \lambda z + L \sin \frac{ \theta}{2} \right) + i \left( f z + L \cos \frac{ \theta }{2} \right) }{ \lambda + i (1-f) } \right) + i 2 \pi k , \quad k \in \Z ,
\end{equation*}
where we defined $L$ and $\theta$ as:
\begin{equation*}
	L \equiv L (\lambda, f) = \sqrt{ 1 + (\lambda^2 + f^2 )^2 \sinh (m)^4 - 2 (\lambda^2 - f^2 ) \sinh(m)^2 } , \quad \sin \theta = \frac{ 2 \lambda f \sinh (m)^2 }{L} , \ \cos \theta = \frac{ 1- (\lambda^2 - f^2) \sinh (m)^2 }{L} .
\end{equation*}
Those saddle points are in general complex, and there is no critical surface in parameter space signalling a phase transition. The sub-critical phase of the case $r=0$ now corresponds to the system living in the surface in the $(\lambda, f , m)$ space determined by the equation:
\begin{equation*}
	\left(  \lambda z + L \sin \frac{ \theta}{2} \right)^2 + \left( f z + L \cos \frac{ \theta }{2} \right)^2 = \lambda^2 + (1-f)^2 ,
\end{equation*}
while the rest of $3d $ parameter space is qualitatively analogous to the super-critical phase of the partition function.

\hypertarget{app:derivatives}{\section{Appendix C. Partition function in the \emph{super-critical phase} for $n=2$}}
The non-Abelian theory with $n=2$ has four relevant saddle points, obtained from the combinations $(\bx_1, \bx_2)=(x_s ^{\pm} , x_s ^{\pm})$. In the \emph{sub-critical phase}, only $(x_s ^{+}, x_s ^{+})$ contributes, but in the \emph{super-critical phase} all four saddle points are to be taken into account, and the partition function is therefore the sum of four pieces:
\begin{equation*}
	\mz^{U(2)}_{\mathrm{super.}} = \mz (x_s ^{+} , x_s^{+}) + \mz (x_s ^{+} , x_s^{-}) + \mz (x_s ^{-} , x_s^{+}) + \mz (x_s ^{-} , x_s^{-}) .
\end{equation*}
Taking advantage of the relations of Appendix \hyperlink{app:derivatives}{A}, one immediately finds:
\begin{equation*}
\mz (x_s ^{+} , x_s^{+}) + \mz (x_s ^{-} , x_s^{-}) = \mz^{U(2)}_{\mathrm{sub.}} + c.c. ,
\end{equation*}
at order $\mathcal{O} (N^{-1})$. The sum of the other two contributions is:
\begin{align*}
	\mz (x_s ^+ , x_s ^- ) + \mz (x_s ^- , x_s ^+ ) & = \frac{\pi e^{- 2 N  \Re  S_1 (x_s ^+)  } }{ 2^{2(N-1)} N^2 } \left\{  \frac{  2 \Re S_1 ^{\prime \prime} (x_s ^+ ) }{ \lvert  S_1 ^{\prime \prime} (x_s ^+) \rvert^{3} } \right. \\
		& + \frac{1}{N} \left. \left[ \frac{ \left( \Re S_1^{\prime \prime} (x_s^+) \right)^2 }{ \lvert  S_1^{\prime \prime} (x_s^+) \rvert^5 }  - \frac{ \Re \left( \left( S_1^{\prime \prime} (x_s ^+) \right)^2 \left( S_1^{(iv)} (x_s^+) \right)^{\ast} \left(5 S_1^{\prime \prime } (x_s ^+) + \left( S_1^{\prime \prime } (x_s ^+ ) \right)^{\ast} \right)  \right)  }{ 4 \lvert  S_1 ^{\prime \prime} (x_s ^+) \rvert^7 } + \right. \right. \\
		& + \left. \left.  \frac{ 5 \Re \left( \left( S_1 ^{\prime \prime} (x_s ^+ ) \right)^3 \left( \left( S_1 ^{\prime \prime \prime} (x_s ^+ ) \right)^{\ast} \right)^2  \left( 7 S_1 ^{\prime \prime} (x_s ^+ ) + \left( S_1 ^{\prime \prime} (x_s ^+ ) \right)^{\ast} \right) \right) - 6  \lvert  S_1 ^{\prime \prime} (x_s ^+ ) \rvert^4  \lvert  S_1 ^{\prime \prime \prime} (x_s^+ ) \rvert^2 }{ 12 \lvert  S_1 ^{\prime \prime} (x_s ^+ ) \rvert^9 }  \right] \right\} .	
\end{align*}

\hypertarget{app:Unc}{\section{Appendix D. Non-Abelian theory: the general case}}
The same procedure applied in the text for the case of $U(2)$ yields in principle for any $U(n)$ theory, \textit{i.e.} arbitrary number of colours, as long as $n$ is kept fixed in the large $N$ limit. At finite $N$, one has the determinantal representation:
\begin{equation*}
	\mz ^{U(n)} _N = N! \det_{1 \le j,k \le N } [ Z_{jk} ] = N! \det_{1 \le j,k \le N } \langle W_{j+k-2} \rangle .
\end{equation*}
Here we compute the large $N$ limit of the partition function \eqref{eq:znc} of the $U(n)$ theory, and the $1/N$ corrections might be obtained in the same fashion as for the $U(2)$ case. The key observation is that, for every $n$, the interaction among eigenvalues is sub-leading as $N \to \infty$, and therefore the set of saddle points of the $U(n)$ theory is given by $n$ copies of the set $\sads$ of the Abelian theory. Another simplification arises from the observation that, at leading order in $1/N$, the determinant is linearized:
\begin{equation*}
	 \prod_{1 \le j < k \le n} \left( 2 \sinh \frac{ x_j - x_k}{2} \right)^2 =  \prod_{1 \le j < k \le n}  \frac{ \left( t_j - t_k  \right)^2 }{N} + \mathcal{O} (N^{-2}) .
\end{equation*}
Consequently, at large $N$ the partition function $\mz_N ^{U(n)}$ converges to:
\begin{equation*}
\label{eq:znc_largeN}
	\mz ^{U(n)} _{\mathrm{sub.}} = \frac{e^{- n N  S_1 (x_s^{+}) } }{2^{n N} N^{\frac{n ^2 }{2}} }  \mz_{\mathrm{GUE} } ( S^{\prime \prime} _1 (x_s ^{+}) ) = \frac{ (2 \pi )^{\frac{n}{2} } e^{- n N  S_1 (x_s^{+}) } }{ 2^{n N} N^{\frac{n ^2 }{2}} ( S^{\prime \prime} _1 (x_s ^{+}) )^{\frac{n ^2 }{2 }} } G (n+2) ,
\end{equation*}
when $\lambda \sinh (m) <1$, where $\mz_{\mathrm{GUE} }(g)$ denotes the partition function of a Gaussian ensemble with coefficient $g$ in the exponent, and $G(n+2) = \prod_{k=0} ^{n} (k!)$ is the Barnes $G$-function. In the \emph{super-critical phase}, $\mz ^{U(n)} _{\mathrm{super.}}$ is a sum over all possible combinations $(\bx_1, \dots, \bx_{n})= ( x_s ^{\pm}, \dots, x_s ^{\pm})$. It is formally given by:
\begin{equation*}
	\mz ^{U(n)} _{\mathrm{super.}} = \frac{ (2 \pi)^{\frac{n}{2}}}{ 2^{n N }  N^{\frac{ n ^2 }{2}} } \sum_{(\bx_1, \dots, \bx_{n} ) \in \sads ^{n} }  \prod_{j=1} ^{n} \frac{ e^{-N  S_1 (\bx_j) } }{ ( S^{\prime \prime} _1 (\bx _j) )^{n - \frac{1 }{2} } } P_{n} \left( S^{\prime \prime} _1 (\bx _1) , \dots, S^{\prime \prime} _1 (\bx _{n} )\right) ,
\end{equation*}
with $P_{n} (s_1, \dots, s_{n})$ a symmetric polynomial of degree $n (n-1)/2$ in $n$ variables, subject to the additional constraint:
\begin{equation*}
	P_{n} (s, \dots, s) = G(n +2) s^{n (n -1) /2 } . 
\end{equation*}
For example, in the $U(3)$ theory it is:
\begin{equation*}
	P_{3} (s_1, s_2, s_3) =  3 \left( s_1 ^2 s_2 + s_1 ^2 s_3 + s_1 s_2^2 +s_1 s_3 ^2 + s_2 ^2 s_3 + s_2 s_3 ^2 - 2 s_1 s_2 s_3 \right) ,
\end{equation*}
and for $U(4)$ it is:
\begin{align*}
	P_{4} (s_1, s_2, s_3, s_4) & = 9 \left\{ 5 s_2 s_3 s_4 \left[ s_2 (s_3 - s_4)^2 + s_2 ^2 (s_3 + s_4) + s_3 s_4 (s_3 + s_4) \right]  \right. \\
	& +  s_1 ^3 \left[ 5 s_2 ^2 (s_3 + s_4) + 5 s_3 s_4 (s_3 + s_4) + s_2 (5 s_3 ^2 - 18 s_3 s_4 + 5 s_4 ^2) \right] \\
	& + s_1^2 \left[ 5 s_3 (s_3 - s_4)^2 s_4 + 5 s_2^3 (s_3 + s_4) - 2 s_2 ^2 (5 s_3 ^2 - 2 s_3 s_4 + 5 s_4 ^2) + s_2 (s_3 + s_4) (5 s_3 ^2 - s_3 s_4 + 5 s_4 ^2) \right] \\
	& \left. + s_1 \left[ 5 s_3 ^2 s_4 ^2 (s_3 + s_4) + s_2 ^3 (5 s_3 ^2 - 18 s_3 s_4 + 5 s_4 ^2) + s_2^2 (s_3 + s_4) (5 s_3^2 - s_3 s_4 + 5 s_4^2) - 2 s_2 s_3 s_4 (9 s_3^2 - 2 s_3 s_4 + 9 s_4^2) \right] \right\} .
\end{align*}
The expression may be further simplified, using the fact that every combination $(\bx_1, \dots, \bx_{n})$ with a fixed number $l$ of entries equal to $x_s ^{+}$, and the remaining $n -l$ equal to $x_s ^{-}$, give the same contribution, independently on the position the $x_s ^{\pm}$ appear. We obtain:
\begin{equation*}
	\mz ^{U(n)} _{\mathrm{super.}} = \frac{ (2 \pi)^{\frac{n}{2}}}{ 2^{n N }  N^{\frac{ n ^2 }{2}} } \sum_{l=0} ^{n} \frac{ e^{-N l S_1 (x_s ^{+}) - N(n -l) S_1 (x_s ^{-}) } }{ (S_1 ^{\prime \prime} (x_s ^{+}) )^{\left( n - \frac{1}{2}\right) l } (S_1 ^{\prime \prime} (x_s ^{-}) )^{\left( n - \frac{1}{2}\right) (n- l) } } \left( \begin{matrix} n  \\ l \end{matrix} \right) P_{n} ( \underbrace{ s , \dots, s  }_{ l} ,  \underbrace{ s^{\ast} , \dots, s ^{\ast} }_{ n - l} ) ,
\end{equation*}
where for shortness we denoted $s \equiv S^{\prime \prime} _1 (x_s ^{+})$ and used $S^{\prime \prime} _1 (x_s ^{-}) = S^{\prime \prime} _1 (x_s ^{+}) ^{\ast} \equiv s^{\ast} $ from Appendix \hyperlink{app:derivatives}{A}.\par
To find the free energy, we reason as in \ccc~for the Abelian case. We write:
\begin{align*}
	\mz ^{U(n)} _{\mathrm{super.}} & \propto \sum_{l=0} ^{n} \exp \left[ -N l S_1 (x_s ^{+}) - N(n -l) S_1 (x_s ^{-})  + \dots \right] \\
	& =  \exp \left[ -n N \Re \left( S_1 (x_s ^{+}) \right) + \log \left( 1+ \sum_{l=0} ^{n} \cos \left( l N \Im \left( S_1 (x_s ^{+}) \right) \right) \right) + \dots \right] ,
\end{align*}
where the dots contain sub-leading terms at large $N$, and arrive to a closed formula for the free energy in the arbitrary $U(n)$ case:
\begin{equation*}
	\mf^{U(n)} = n \mf ^{U(1)} .
\end{equation*}

\hypertarget{app:Rcharge}{\section{Appendix E. General $R$-charges}}
The partition function of the $U(1)$ $\mn =2$ theory with $N$ chiral multiplets of mass $m$ and $N$ chiral multiplets of mass $-m$ with arbitrary $R$-charge $q$, and coupled to a FI background, is \cite{Jafferis:2010un,Hama:2010av}:
\begin{align*}
	\mz_{Z,q} ^{U(1)} = \int_{- \infty } ^{\infty} dx \exp   \big\{  i \eta x   & + N \left[ \ell \left( 1 - q + \frac{i (x+m)}{2 \pi} \right) + \ell \left( 1 - q - \frac{i (x+m)}{2 \pi} \right) \right]  \\
	& \left. + N \left[ \ell \left( 1 - q + \frac{i (x-m)}{2 \pi} \right) + \ell \left( 1 - q - \frac{i (x-m)}{2 \pi} \right) \right] \right\} ,
\end{align*}
where we recall that the theory is put on a three-sphere of radius $1/2 \pi$. Here, $e^{\ell (z)}$ is the double sine function, defined as \cite{Jafferis:2010un}:
\begin{equation*}
    \ell (u) = -u \log \left( 1 - e^{i 2 \pi u} \right) + \frac{i \pi}{2} u^2 + \frac{i}{2 \pi} \mathrm{Li}_2 \left( e^{i 2 \pi u } \right) - \frac{i\pi}{12} , \qquad u \in \C .
\end{equation*}
This function has logarithmic singularities when $q \in \Z$, or, more in general, when $\Im \left( \widetilde{m} \right) + 2 \pi (1-q) \in 2 \pi \Z$, where $\widetilde{m}$ denotes a complexified mass parameter. Nevertheless, the partition function does not develop singularities, and in fact is holomorphic in $\widetilde{m}$, as the divergences cancel. This can be seen, for instance, from the identity
\begin{equation*}
	\ell \left( 1-q - \frac{ i(x-m)}{2 \pi} \right) \stackrel{\mathrm{reg.}}{=} - \ell \left( 1-q + \frac{ i(x- \widetilde{m})}{2 \pi} \right) , \qquad \widetilde{m} = m - i 4 \pi (1-q) ,
\end{equation*}
where the equality is exact for the infinite product representation of the one-loop determinants and extends to the function $\ell$ through regularization by $\zeta$-function.

The derivative of the double sine function satisfies the simple property:
\begin{equation*}
	\frac{ d \ell }{ du} = - \pi u \cot (\pi u ) .
\end{equation*}
Therefore, in the double scaling large $N$ limit, we arrive to the saddle point equation:
\begin{equation}
\label{eq:speq}
	\frac{  \frac{(x+m)}{2 \pi} \sin \left( 2 \pi (1-q) \right) - i (1-q) \sinh (x+m) }{ \cosh (x+m) - \cos \left( 2 \pi (1-q) \right) } +  \frac{  \frac{(x-m)}{2 \pi} \sin \left( 2 \pi (1-q) \right) - i (1-q) \sinh (x-m) }{ \cosh (x-m) - \cos \left( 2 \pi (1-q) \right) } = \lambda .
\end{equation}
It is a simple exercise to see that, setting $q = \frac{1}{2}$, one recovers the saddle points of the $\mn =4$ theory \ccc.~When $q$ is half-integer, the trigonometric functions take simple values and we can solve the saddle point equation exactly, as showed in the main text.\par
\begin{figure}[bth]
\centering
	\includegraphics[width=0.5\textwidth]{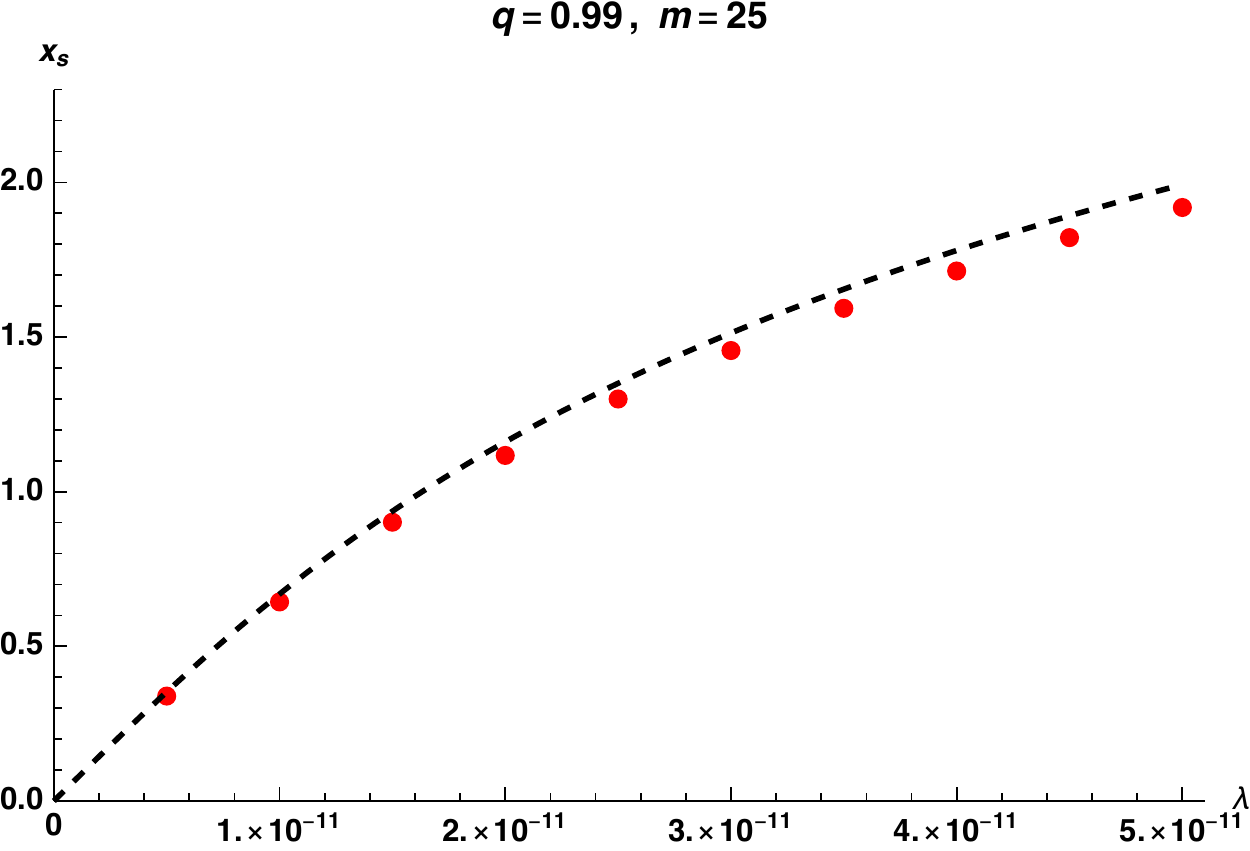}
	\caption{Comparison of the numerical solution of the saddle point equation \eqref{eq:speq} (red dots) and expression \eqref{eq:approxqlargem} (black, dashed line), for $q= 0.99$ and $m=25$.}
\label{fig:numericq}
\end{figure}

We found out that, for $q=1$, the action admits no saddle point. Here, we study what happens close to that point, for real $q=1-\varepsilon$. We assume $\varepsilon$ small and approximate the expression at $\mathcal{O} (\varepsilon)$. From \eqref{eq:speq} we get:
\begin{equation*}
	\frac{ \sinh (x) }{ \cosh (x) - \cosh (m) } + i \frac{ x ( \cosh (x) \cosh (m) -1 ) + m \sinh (x) \sinh (m) }{ \left( \cosh (x) - \cosh (m)  \right)^2 } = \frac{ i \lambda }{2 \varepsilon} .
\end{equation*}
The equation is still transcendental, but we can find an approximate solution in the large mass limit:
\begin{equation}
\label{eq:approxqlargem}
	\sinh (x) \approx \frac{ \lambda  e^{m} }{2 m \varepsilon }  \ \Longrightarrow \ x \approx \log \left[ \frac{ \lambda e^{m} }{2 m \varepsilon } \left( 1 + \sqrt{ 1 + \frac{4 m^2 \varepsilon^2 e^{-2m} }{ \lambda^2} } \right) \right] \approx m + \log \frac{ \lambda }{m \varepsilon } .
\end{equation}
In figure \ref{fig:numericq} we compare this expression at large $m$ with a numerical solution to the saddle point equation.

\subsection*{Comment on squashed geometry}
If the supersymmetric $\mn = 2$ theory is put on a squashed three-sphere instead than a round one, the partition function is obtained replacing the double sine functions by their squashed version \cite{Hama:2011ea}
\begin{equation*}
	\ell \left( 1-q + \frac{i \sigma}{2 \pi} \right) \ell \left( 1-q - \frac{i \sigma}{2 \pi} \right) \mapsto \ell_b \left( \frac{1}{2} \left( b + \frac{1}{b} \right) (1-q) + \frac{i \sigma}{2 \pi} \right) \ell_b \left( \frac{1}{2} \left( b + \frac{1}{b} \right) (1-q) - \frac{i \sigma}{2 \pi} \right), \qquad \sigma = x \pm m ,
\end{equation*}
where $b= \sqrt{r_1/r_2}$ is the squashing parameter, and the average radius is $\sqrt{r_1 r_2} = 1/ 2 \pi$. We now take advantage of the remarkable property of the double sine function:
\begin{equation*}
	\exp \left\{ \ell_b \left( \frac{b}{2} + \frac{i \sigma}{2 \pi} \right) \ell_b \left( \frac{b}{2} - \frac{i \sigma}{2 \pi} \right) \right\} = \frac{1}{2 \cosh \left( \frac{ b \sigma }{2} \right) } ,
\end{equation*}
which holds for every real non-negative $b$, and for the round case $b=1$ provides the partition function \eqref{eq:z1}. Therefore, for hypermultiplets with $R$-charge $0<q<1$, we may tune the geometry of the manifold so that $(b+b^{-1})(1-q)=b$, that is we may squash the sphere as
\begin{equation*}
	b = \sqrt{ \frac{ 1-q}{q} } ,
\end{equation*}
and the partition function reads:
\begin{equation*}
	\mz ^{U(1)} _{N, \mathrm{squash}} = \int_{- \infty} ^{\infty} dx \frac{ e^{i \eta x } }{ 2^N \left[ \cosh (bx) + \cosh (bm) \right]^N } .
\end{equation*}
Notice that this procedure would also affect the one-loop determinant of the vector multiplet, but this is irrelevant in the Abelian theory, being such determinant trivial. Also, we see that the symmetry $q \leftrightarrow 1-q$ at the matrix model level is translated into a symmetry $b \leftrightarrow b^{-1}$ in the geometry. We therefore obtain a simple relation between the partition function of the $\mn =2$ theory with arbitrary $R$-charge $0<q<1$ posed on a suitably squashed sphere and the $\mn =4$ theory with $R$-charge $q=\frac{1}{2}$ on the round $\cS^3$:
\begin{equation*}
	\mz ^{U(1)} _{N, \mathrm{squash}} (m, \eta, q) = \frac{1}{b} \mz ^{U(1)} _{N, \mathrm{round}} \left( b m, \frac{ \eta}{b} , q= \frac{1}{2} \right) , \qquad b= \sqrt{\frac{1-q}{q}} .
\end{equation*}
As a byproduct, this equivalence holds for the $U(2)$ theory in the large $N$ approximation. In fact, the squashing would modify:
\begin{equation*}
	\left( \sinh \frac{ x_1 - x_2}{2} \right)^2 \mapsto \left( \sinh \frac{ b(x_1 - x_2)}{2} \right) \left( \sinh \frac{ x_1 - x_2}{2 b} \right) ,
\end{equation*}
and, as we have seen, the determinant is linearized at first order in $1/N$, producing cancellation of the $b$-dependence.

\end{appendix}

\end{document}